\documentclass[prd,aps,floats,twocolumn,twoside,preprintnumbers,
superscriptaddress,floatfix,nofootinbib,showpacs]{revtex4}
\usepackage{graphicx,pstricks,rotating,amsmath}
\usepackage{amssymb,bbm}
\usepackage{slashed}
\begin{document} 


\title{Equivalence of quantum field theories related by the $\theta$-exact Seiberg-Witten map}

\author{Carmelo P. Martin}
\affiliation{Departamento de F\'{\i}sica Te\'orica I, Facultad de Ciencias F\'{\i}sicas,
Universidad Complutense de Madrid, 28040-Madrid, Spain}
\email{carmelop@fis.ucm.es}
\author{Josip Trampeti\'{c}}
\affiliation{Institute Rudjer Bo\v{s}kovi\'{c}, Division of Experimental Physics, Bijeni\v{c}ka 54, 10000 Zagreb, Croatia}
\email{josip@irb.hr}
\affiliation{Max-Planck-Institut f\"ur Physik, (Werner-Heisenberg-Institut), F\"ohringer Ring 6, D-80805 M\"unchen, Germany}
\email{trampeti@mppmu.mpg.de}
\author{Jiangyang You}
\affiliation{Institute Rudjer Bo\v{s}kovi\'{c}, Division of Theoretical Physics, Bijeni\v{c}ka 54 10000 Zagreb, Croatia}
\email{youjiangyang@gmail.com}

\newcommand{\tr}{\hbox{tr}}
\def\BOX{\mathord{\vbox{\hrule\hbox{\vrule\hskip 3pt\vbox{\vskip
3pt\vskip 3pt}\hskip 3pt\vrule}\hrule}\hskip 1pt}}

\date{\today}

\begin{abstract}
The equivalence of the noncommutative U(N) quantum field theories related by the $\theta$-exact Seiberg-Witten maps is, in this paper, proven to all orders in the perturbation theory with respect to the coupling constant. We show that this holds for super Yang-Mills theories with ${\cal N}=0,1,2,4$ supersymmetry. A direct check of this equivalence relation is  performed by computing the one-loop quantum corrections to the quadratic part of the effective action in the noncommutative $\rm U(1)$ gauge theory with ${\cal N}=0,1,2,4$ supersymmetry.
\end{abstract}


\maketitle
Classical noncommutative (NC) field theories admit an equivalent representation in terms of ordinary fields formulated by employing the Seiberg-Witten (SW) map~\cite{Seiberg:1999vs}. However, we still do not know whether this equivalence holds at the quantum level, i.e., whether the quantum theory defined in terms of the NC fields is the same as the quantum theory defined in terms of commutative fields and obtained from the NC action by using the $\theta$-exact SW map~\cite{Seiberg:1999vs}. In this communication, we prove that the $\theta$-exact  Seiberg-Witten map establishes an equivalence relation between perturbative--in the coupling constant--quantum field theories defined with respect to the noncommutative and commutative fields, by showing that the corresponding on-shell DeWitt effective actions~\cite{DeWitt:1967ub,Kallosh:1974yh,DeWitt:1980jv,DeWitt:1988fm} can be SW-mapped one to another. We also give an explicit check of our verdict in the (supersymmetric) NC $\rm U(1)$ gauge theory.

The on-shell DeWitt effective action \cite{DeWitt:1988fm} with respect to the noncommutative or hatted fields, $\hat\Gamma_{\rm DeW}\big[\hat B_\mu\big]$, is given by the following path integral formulation,
\begin{equation}
\begin{split}
&e^{\frac{i}{\hbar}\hat\Gamma_{\rm DeW}\big[\hat B_\mu\big]}=\int d\hat Q_\mu^a d\hat C^a d\hat{\bar C}^a d\hat F^a\;
\\&\cdot e^{\frac{i}{\hbar}S_{\rm NCYM}\big[\hat B_\mu+\hbar^{\frac{1}{2}}\hat Q_\mu\big]+i S_{\rm gf}\big[\hat B_\mu,\hat Q_\mu,\hat F,\hat{\bar C},\hat C\big]},
\end{split}
\label{effectivenoncommutative}
\end{equation}
where $S_{\rm NCYM}=-\frac{1}{4g^2}\int \tr\,\hat F_{\mu\nu}\hat F^{\mu\nu}$ is the usual NC $\rm U_\star(N)$ Yang-Mills (YM) action, while $S_{\rm gf}$ is the gauge-fixing action which can be expressed in the Becchi-Rouet-Stora-Tyutin (BRST) quantization language as
\begin{equation}
S_{\rm gf}\big[\hat B_\mu,\hat Q_\mu,\hat F,\hat{\Bar C},\hat C\big]=\hat\delta_{BRS}\,X_{\rm gf}\big[\hat B_\mu,\hat Q_\mu,\hat F,\hat{\Bar C},\hat C\big],
\label{Xgf}
\end{equation}
where $X_{\rm gf}\big[\hat B_\mu,\hat Q_\mu,\hat F,\hat{\Bar C},\hat C\big]$ is an arbitrary gauge-fixing functional. The noncommutative 
$\rm U_\star(N)$ BRS transformations $\hat\delta_{BRS}\hat A_\mu=\hat D_\mu\hat C$ and $\hat\delta_{BRS}\hat C=-i\hat C\star\hat C$ induce the following BRS transformations after 
background-field splitting
$\hat A_\mu\to \hat B_\mu+\hbar^{\frac{1}{2}}\hat Q_\mu$,

\begin{equation}
\begin{array}{cr}
\hat\delta_{\rm BRS}\hat B_\mu=0,\;\hat\delta_{\rm BRS}\hat Q_\mu=\hbar^{-\frac{1}{2}} \hat D_\mu\big[\hat B_\mu+\hbar^{\frac{1}{2}}\hat Q_\mu\big]\hat C,
\\
\hat\delta_{\rm BRS}\hat C=-i\hat C\star\hat C,\;
\hat\delta_{\rm BRS}\hat{\bar C}=\hbar^{-\frac{1}{2}}\hat F,\;
\hat\delta_{\rm BRS}\hat F=0.
\end{array}
\label{hatBRS}
\end{equation}

The $\theta$-exact SW map of the NC fields in terms of commutative and ordinary fields in the $\rm U(N)$ gauge theory,
\begin{gather}
\hat A_\mu=\hat A_\mu\left[A_\mu,\theta\right],\:\;
\hat C=\hat C\left[A_\mu,C,\theta\right],
\label{SW}
\end{gather}
are solutions to the following equations:
\begin{gather}
\hat\delta_{\rm BRS}\hat A_\mu=\delta_{\rm BRS}\hat A_\mu\left[A_\mu,\theta\right],\:\:
\hat\delta_{\rm BRS}\hat C=\delta_{\rm BRS}\hat C\left[A_\mu,C,\theta\right].
\label{SWeqs}
\end{gather}
They can be expressed $\theta$-exactly as a formal power series of the field operators~\cite{Martin:2012aw, Martin:2015nna},
\begin{gather}
\hat A_\mu\left[A_\mu,\theta\right](x)=A_\mu(x)+\sum\limits_{n=2}^\infty \mathcal{A}_\mu^{(n)}(x),
\label{SW1}
\\
\hat C\left[A_\mu,C,\theta\right](x)=C(x)+\sum\limits_{n=1}^\infty \mathcal{C}^{(n)}(x),
\label{SW2}
\end{gather}
where
\begin{gather}
\begin{split}
\mathcal{A}_\mu^{(n)}(x)=\int&\prod\limits_{i=1}^n\frac{d^4 p_i}{(2\pi)^4} e^{i\left(\sum\limits_{i=1}^n p_i\right)x}
\\&\cdot\mathfrak{A}^{(n)}_\mu\big[(a_1,\mu_1,p_1),......,(a_n,\mu_n,p_n);\theta\big]
\\&\cdot\tilde A_{\mu_1}^{a_1}(p_1)......\tilde A_{\mu_n}^{a_n}(p_n),
\end{split}
\label{SW3}
\\
\begin{split}
\mathcal{C}^{(n)}(x)&=\int\prod\limits_{i=1}^n\frac{d^4 p_i}{(2\pi)^4} e^{i\left(p+\sum\limits_{i=1}^n p_i\right)x}
\\&\cdot\mathfrak{C}^{(n)}\big[(a_1,\mu_1,p_1),......,(a_n,\mu_n,p_n);(a,p);\theta\big]
\\&\cdot \tilde A_{\mu_1}^{a_1}(p_1)......\tilde A_{\mu_n}^{a_n}(p_n)C^a(p).
\end{split}
\label{SW4}
\end{gather}
The quantities $\mathfrak{A}_\mu^{(n)}$ and $\mathfrak{C}^{(n)}$ are totally symmetric under the permutations with respect to the set of the parameter-triples $\left\{(a_i,\mu_i,p_i)|i=1,...,n\right\}$, which have the property--of key importance--that only the momenta which are not contracted with $\theta^{\mu\nu}$ build up polynomials, which  never occur in the denominator \cite{Martin:2012aw, Martin:2015nna}. 

Introducing the ordinary background-field splitting,
\begin{equation}
A_\mu=B_\mu+\hbar^{\frac{1}{2}}Q_\mu,
\label{ordinarysplit}
\end{equation}
and the corresponding BRS transformations,
\begin{equation}
\delta_{\rm BRS} B_\mu=0,\; \delta_{\rm BRS} Q_\mu=\hbar^{-\frac{1}{2}} D_\mu\big[B_\mu+\hbar^\frac{1}{2} Q_\mu\big]C,
\label{normalBRS}
\end{equation}
where $B_\mu$ is the commutative background field and $Q_\mu$ the commutative quantum fluctuation,  for the SW map \eqref{SW} we find the background-field splitting
\begin{equation}
\begin{split}
\hat A_\mu\big[B_\mu+\hbar^{\frac{1}{2}}Q_\mu,\theta\big]=&\hat A_\mu\big[B_\mu,\theta\big]+\hbar^{\frac{1}{2}}\hat Q_\mu\big[B_\mu,Q_\mu,\hbar,\theta\big]
\\=&\hat B_\mu\big[B_\mu,\theta\big]+\hbar^{\frac{1}{2}}\hat Q_\mu\big[B_\mu,Q_\mu,\hbar,\theta\big]
\\
\hat C\big[B_\mu+\hbar^{\frac{1}{2}}Q_\mu,C,\theta\big]&=\hat C\big[B_\mu,C,\theta\big]
\\&+\hbar^{\frac{1}{2}}\hat C^{(1)}\big[B_\mu,Q_\mu,C,\hbar,\theta\big],
\end{split}
\label{Qdefinition}
\end{equation}
which ensures that the ordinary BRS transformations \eqref{normalBRS} induce the NC BRS transformations \eqref{hatBRS}.

Now the on-shell DeWitt action with respect to the ordinary fields, $\Gamma_{\rm DeW}\left[B_\mu\right]$, is given by the path integral
\begin{equation}
\begin{split}
&e^{\frac{i}{\hbar}\Gamma_{\rm DeW}\big[B_\mu\big]}=\int d Q_\mu^a d C^a d \hat{\bar {C^a}} d \hat F^a\;
\\&
\cdot e^{\frac{i}{\hbar}S_{\rm NCYM}\big[B_\mu+\hbar^{\frac{1}{2}}Q_\mu\big]+i  S_{\rm gf}\big[ B_\mu, Q_\mu,\hat F,\hat{\bar C},C\big]},
\end{split}
\label{effectiveaction}
\end{equation}
in which we change variables: $C^a \to \hat C^a$ and $Q^a\to\hat Q^a$, so that it transforms into the new path integral
\begin{eqnarray}
&&{e^{\frac{i}{\hbar}\Gamma_{\rm DeW}\big[B_\mu\big]}=\int d\hat Q_\mu^a d \hat C^a d\hat{\bar C}^a d \hat F^a\;J^{-1}_1[B,Q]\,J_2[B,Q]\,}\nonumber\\
&&{\cdot e^{\frac{i}{\hbar}S_{\rm NCYM}\big[\hat B_\mu+\hbar^{\frac{1}{2}}\hat Q_\mu\big]+i  S_{\rm gf}\big[\hat B_\mu, \hat Q_\mu,\hat F,\hat{\bar C},\hat C\big]},}
\label{effectiveactionchanged}
\end{eqnarray}
containing the Jacobian determinants $J_1\big[B^a,Q^a\big]$ and $J_2\big[B^a,Q^a\big]$ which are defined as follows:
\begin{equation}
\begin{array}{l}
{J_1\big[B^a, Q^a\big]\,=\,\det\frac{\delta\hat Q^a_\mu(x)}{\delta Q^b_\nu(y)}\,=\,\exp\,{\rm Tr}\,\ln\Big(\frac{\delta\hat Q^a_\mu(x)}{\delta Q^b_\nu(y)}\Big)},\\[8pt]
{ J_2\big[B^a,Q^a\big]\,=\,\det\frac{\delta\hat C^a(x)}{\delta C^b(y)}\,=\,\exp\,{\rm Tr}\,\ln\Big(\frac{\delta\hat C^a(x)}{\delta C^b(y)}\Big).}
 \end{array}
 \label{jacobdet}
 \end{equation}

Under the assumption that both of the above Jacobians are equal to one,
we can prove that the right-hand side of (\ref{effectiveactionchanged}) equals to the right-hand side of (\ref{effectivenoncommutative}), so that
\begin{equation}
\Gamma_{\rm DeW}\big[B_\mu\big]=\hat\Gamma_{\rm DeW}\big[\hat B_\mu[B_\mu]\big].
\label{equiv}
\end{equation}

Note that the above result is valid on-shell, i.e. when $\hat B_\mu[B_\mu]$ satisfies the NC YM equations of motions 
\begin{equation}
\hat D_{\mu}\big[\hat B_\mu[B_\mu]\big]\hat F^{\mu\nu}\big[\hat B_\mu[B_\mu]\big]=0,
\label{NCYMEOM}
\end{equation}
and the reason is the on-shell uniqueness of the DeWitt effective action~\cite{Kallosh:1974yh,Ichinose:1992np}.

Using the SW map expansion \eqref{SW3} and the background-field splitting (\ref{Qdefinition}) one can show that
\begin{equation}
\begin{array}{l}
{\frac{\delta\hat Q^a_\mu(x)}{\delta Q^b_\nu(y)}\,=\,\frac{1}{\hbar^{\frac{1}{2}}}\,\frac{\delta\hat A^a_\mu(x)}{\delta Q^b_\nu(y)}
=\delta^a_b\delta^\nu_\mu\,\delta(x-y)
+\sum\limits_{n=2}^{\infty}\,
\int\prod\limits_{i=1}^n\frac{d^4 p_i}{(2\pi)^4}}
\\[8pt]{\cdot
e^{i\left(\sum\limits_{i=1}^{n-1} p_i\right)x}\,e^{ip_n (x-y)}
{\cal M}^{(n)\,a\,\nu}_{\phantom{(n)\,}b\,\mu}(p_1,p_2,....p_{n-1};p_n;\theta)},
\end{array}
\label{dethQQ}
\end{equation}
where
\begin{equation}
\begin{split}
&{\cal M}^{(n)\,a\,\nu}_{\phantom{(n)\,}b\,\mu}(p_1,p_2,....p_{n-1};p_n;\theta)
\\
&=n\;\tr\Big[T^a{\mathfrak{A}^{(n)}_\mu\big[(a_1,\mu_1,p_1)},
...,(a_{n-1},\mu_{n-1},p_{n-1}),
\\&
(b,\nu,p_n);\theta\big]\Big]
\tilde A_{\mu_1}^{a_1}(p_1)...\tilde A_{\mu_{n-1}}^{a_{n-1}}(p_{n-1}).
\end{split}
\label{Mcaldef}
\end{equation}
Note  that $\tilde A_{\mu_i}^{a_i}(p_i)=\tilde B_{\mu_i}^{a_i}(p_i)+\hbar^{\frac{1}{2}}\tilde Q_{\mu_i}^{a_i}(p_i)$ for all $i$.

Let $l_i$, $i=1,..,m+1$ be given by
\begin{equation*}
l_1=\sum\limits_{i_1=1}^{n_1-1}\, p_{1,i_1}, 
\:...........\:, l_{m+1}=\sum\limits_{i_{m+1}=1}^{n_{m+1}}\,p_{m+1,i_{m+1}},
\end{equation*}
then, by taking into account (\ref{dethQQ}) and carrying out a lengthy straightforward computation, one gets
\begin{equation}
\begin{array}{l}
{\ln\,J_1[B,Q]={\rm Tr}\ln\,\Big(\frac{\delta\hat Q^a_\mu(x)}{\delta Q^b_\nu(y)}\Big)={\sum\limits_{n=2}^{\infty}}
{\int\prod\limits_{i=1}^{n-1}\frac{d^4 p_i}{(2\pi)^4}\,\delta\Big(\sum\limits_{i=1}^{n-1}p_i\Big)}}
\\[8pt]
{\cdot\int\frac{d^4 q}{(2\pi)^4}\,
{\cal M}^{(n)\,a\,\mu}_{\phantom{(n)\,}a\,\mu}\left(p_1,p_2,....,p_{n-1};q;\theta\right)+\sum\limits_{m=1}^{\infty}\frac{(-1)^{m}}{m+1}}
\\[8pt]
{\cdot\sum\limits_{n_1=2}^{\infty}\cdots\sum\limits_{n_{m+1}
=2}^{\infty}
\int\prod\limits_{i_1=1}^{n_1-1}\frac{d^4 p_{1,i_1}}{(2\pi)^4}\cdots\int\prod\limits_{i_{m+1}=1}^{n_{m+1}-1}\frac{d^4 p_{m+1,i_{m+1}}}{(2\pi)^4}}
\\[8pt]
{\cdot\delta\Big(\sum\limits_{i=1}^{m+1}l_i\Big)\int\frac{d^4 q}{(2\pi)^4}\Big[
{\cal M}^{(n_1)\,a\,\mu_1}_{\phantom{(n_1)\,}a_1\,\mu}\left(p_{1,1},p_{1,2},....,p_{1,n_1-1};q;\theta\right)
}\\[8pt]
{\cdot{\cal M}^{(n_2)\,a_1\,\mu_2}_{\phantom{(n_3)\,}a_2\,\mu_1}\left(p_{2,1},p_{2,2},....,p_{2,n_2-1};q-l_2;\theta\right)}
\\[8pt]
{\cdots\cdots\cdots}
\\[8pt]
{\cdot{\cal M}^{(n_{m+1})\,a_m\,\mu}_{\phantom{(n_{m+1}\,}a\,\,\,\mu_m}\Big(p_{m+1,1},p_{m+1,2},
....,p_{m+1, n_{m+1}-1,}}
\\[8pt]
{q-\sum\limits_{i=2}^{m+1}l_i;\theta\Big)\Big].}
\label{genexpln}
\end{array}
\end{equation}
Hence, in view of the above equations (\ref{Mcaldef}) and (\ref{genexpln}), to compute $\ln\,J_1[B,Q]$, one has to work out the following dimensionally regularized type of integrals over the internal momenta $q^\mu$:
\begin{equation}
\begin{array}{l}
{\mathfrak{V}=\int\frac{d^D q}{(2\pi)^D}\Big\{
{\tr\Big[T^a{\mathfrak{A}_\mu^{(n_1)}}\big[(b_{1,1},\nu_{1,1},p_{1,1}),}}
\\[8pt]
{.....,(b_{1,n_1-1},\nu_{1,n_1-1},p_{1,n_1-1}),(a_1,\mu_1,q);\theta\big]\Big]}
\\[8pt]
{\cdots\cdots\cdots}
\\[8pt]
{\cdot\tr\Big[T^{a_m}{\mathfrak{A}^{(n_{m+1})}}_{\mu_m}\big[(b_{m+1,1},\nu_{m+1,1},p_{m+1,1}),....,}
\\[8pt]
{(b_{m+1,n_{m+1}-1},\nu_{m+1,n_{m+1}-1},p_{m+1,n_{m+1}-1}),}
\\[8pt]
{(a,\mu,q-\sum\limits_{i=2}^{m+1} l_i);\theta\big]\Big]\Big\}.}
\label{dimregint}
\end{array}
\end{equation}
However, the previous integral in (\ref{dimregint}) is a linear combination of integrals of the type
\begin{equation}
\mathfrak{I}\,=\,\int\frac{d^D q}{(2\pi)^D}\,\mathbb{Q}(q)\,\mathbb{I}(q\theta k_i,k_i\theta k_j),
\label{mathI}
\end{equation}
where $\mathbb{Q}(q)=q^{\rho_1}q^{\rho_2}q^{\rho_3}\cdots $, $q\theta k_i=q_\mu\theta^{\mu\nu}k_{i\nu}$, 
and $k_i\theta k_j=k_{i\mu}\theta^{\mu\nu}k_{j\nu}$.  Indices $i$ and $j$ run over all relevant (external) momenta other than $q$.
It is important to stress that $\mathbb{Q}(q)$ is a monomial on $q^{\rho}$ and that the functional $\mathbb{I}$, as indicated in the integrand of the integral (\ref{mathI}), is a function of the variables $q\theta k_i$ and $k_i\theta k_j$ only, and, hence, as shown in details in \cite{Martin:2016saw}, one concludes that
\begin{equation}
\mathfrak{I}=0\: \,\,\to\:\,\, \mathfrak{V}=0,
\label{22}
\end{equation}
under dimensional regularization \cite{Collins:1984xc}.
By substituting $\mathfrak{V}=0$ in (\ref{genexpln}), we obtain that in dimensional regularization the following result holds,
\begin{equation}
\ln\,J_1[B,Q]=0,
\label{dimregJ1}
\end{equation}
proving that indeed $J_1[B,Q]=1$.

It is straightforward to see that identical arguments apply to $J_2[B,Q]$ as well; thus, the Seiberg-Witten map equivalence between quantum theories defined in terms of noncommutative fields and in terms of ordinary fields indeed holds up to all orders in the perturbation theory.

 Now, since the $\theta$-exact Seiberg-Witten map for matter fields--see \cite{Martin:2012aw, Martin:2015nna}--have expressions analogous to that of the ghost field, it is clear that the Jacobian of the transformation from ordinary matter fields to noncommutative matter fields is also trivial in dimensional regularization. Hence, the conclusion that we have reached above, when no matter fields are included, remains valid when the latter are included: the on-shell De Witt action of the theory defined in terms of noncommutative fields is the same as the on-shell DeWitt action of the ordinary theory obtained by using the $\theta$-exact Seiberg-Witten map.

We have checked the  equivalence established above by computing the one-loop quantum correction to the quadratic part of the effective action of the U(1) NCGFT in the NC background-field gauge prior to and after the Seiberg-Witten map. In this specific case the general equivalence reduces to a simple relation:
\begin{equation}
\hat\Gamma^{\mu\nu}(p)=\Gamma^{\mu\nu}(p)\Big|_{\rm on-shell}.
\label{on-shell}
\end{equation}

The standard procedure for computing the DeWitt effective action of the $\rm U_\star(1)$ gauge theory perturbatively in the background-field formalism~\cite{Kallosh:1974yh,DeWitt:1980jv} evaluates 1-PI diagrams with all background-field external legs and all integrand field ($\hat Q_\mu,\hat{\bar C},\hat C,\hat F$) internal lines using the following action: 
\begin{equation}
\begin{split}
\hat S_{\rm loop}=&S_{\rm BFG}+S_{\rm NCYM}\big[\hat B_\mu+\hat Q_\mu\big]-S_{\rm NCYM}\big[\hat B_\mu\big]
\\&-\int \bigg(\frac{\delta}{\delta \hat B_\mu}S_{\rm NCYM}\big[\hat B_\mu\big]\bigg)\cdot\hat Q_\mu.
\end{split}
\label{NCBFG}
\end{equation}
We choose the $\theta$-exact SW map from $\hat S_{\rm loop}\to S_{\rm loop}$ and then use the resulting action
\begin{equation}
\begin{split}
&S_{\rm loop}=S_{\rm BFG}\big[B_\mu,Q_\mu,\hat{\bar C}, C,\hat F\big]
\\&+S_{\rm NCYM}\Big[\hat B_\mu\big[B_\mu\big]+\hat Q_\mu\big[Q_\mu,B_\mu\big]\Big]
-S_{\rm NCYM}\Big[\hat B_\mu\big[B_\mu\big]\Big]
\\&-\int \bigg(\frac{\delta }{\delta \hat B_\mu}S_{\rm NCYM}\big[\hat B_\mu\big]\bigg)\big[B_\mu\big]\,\cdot\,\hat Q_\mu\big[B_\mu,Q_\mu\big],
\end{split}
\label{SWBFG}
\end{equation}
for the one-loop computation of the effective action with respect to the ordinary fields. This choice can be shown to be equivalent to the subtraction of commutative equations of motion
$\frac{\delta}{\delta B_\mu}S_{\rm NCYM}\big[\hat B_\mu[B_\mu]\big]=0$ on shell as long as the Seiberg-Witten map is invertible.

In the follow-on computation, by using the extended version of the dimensional regularization scheme~\cite{Martin:2016zon}, we find that the one-loop 1-PI two-point functions from \eqref{NCBFG} and \eqref{SWBFG} are actually exactly the same, i.e.\footnote{Explicit computations of the photon polarization tensor $\Gamma^{\mu\nu}$ with full technical details are presented in~\cite{Martin:2016saw}.}
\begin{equation}
\hat\Gamma^{\mu\nu}(p)=\Gamma^{\mu\nu}(p),
\label{hatGamma}
\end{equation}
which verifies the equivalence relation \eqref{equiv}. 

 As a consequence of (\ref{hatGamma}), an important point is that, once we turn on supersymmetry~\cite{Martin:2016zon}, both the IR and UV cancellation results, 
\begin{gather}
\;\;\,\Gamma_{\rm total}^{\mu\nu}\big|_{\rm IR}=\frac{g^2}{\pi^2}\Big(2-2 n_f +  n_s\Big)\frac{(\theta p)^\mu(\theta p)^\nu}{(\theta p)^4},
\label{N=1,2,4sumIR}
\\ 
\begin{split}
\Gamma^{\mu\nu}_{\rm BFGtotal} \big|_{\rm UV}&=\frac{g^2}{48\pi^2}\big(22-4{\rm n_f}-{\rm n_s}\big)
\\&\cdot\big(g^{\mu\nu}p^2-p^\mu p^\nu\big)
\Big(\frac{2}{\epsilon}+\ln(\mu^2(\theta p)^2)\Big),
\label{N=1,2,4sumUV}
\end{split}
\end{gather}
found prior to the Seiberg-Witten map now hold precisely after the Seiberg-Witten map.

A summary of our communication is as follows:

The perturbative quantum field theories derived from the classical action in terms of the noncommutative or the ordinary commutative fields via the Seiberg-Witten map are equivalent to each other, again via the Seiberg-Witten map. This is because each order of the perturbative expansion of the Jacobian determinant associated with the Seiberg-Witten map changing variable contains a single functional trace which can be converted into a single loop integral that vanishes in the dimensional regularization. Therefore, the defining path integral for the on-shell DeWitt effective action becomes identical after changing the path integral variables. From this viewpoint, at least in the perturbative regime, the overwhelming nonlocality which spreads all over the $\theta$-exact  Seiberg-Witten map expansion, together with all higher-order interactions induced by the Seiberg-Witten map, can actually both be minimized to the same level as in the much simpler theory without the Seiberg-Witten map. 

We have checked the general equivalence relation in the one-loop corrections to the quadratic part of both DeWitt effective actions. We find that the equivalence relation manifests itself as an equality between the effective actions when we perform a subtraction of the noncommutative equations of motions. Since the equations of motions with respect to ordinary and noncommutative fields are equivalent as long as the Seiberg-Witten map is invertible, we conclude that the equivalence relation indeed holds on-shell.

The ``fingerprint'' quantum properties of the supersymmetric noncommutative $\rm U(1)$ gauge theories~\cite{Jack:2001cr,Zanon:2000nq,Santambrogio:2000rs,Pernici:2000va,Buchbinder:2001at,Ruiz:2000hu,Ferrari:2003vs,AlvarezGaume:2003mb,Ferrari:2004ex}, namely the cancellation of quadratic IR divergence in the photon 1PI two-point function by SUSY and the cancellation of the UV plus log-IR divergences in the photon 1PI two-point function in the background field gauge by $\mathcal N=4$ SUSY, which is only valid when quantization is performed with respect to the noncommutative fields before, are now also obtainable from the quantization with respect to the ordinary fields.

A few final words on the scope and limits of our results: We require the preexistence of a self-consistent NCGFT which closes on the $\rm U(N)$ Lie algebra without the Seiberg-Witten  map, which admits a sound perturbative quantization by itself and, more importantly, an invertible Seiberg-Witten map. There exists, for example, deformed $\rm SU(N)$ gauge theories~\cite{Jurco:2000ja} possessing a noncommuatative theory which closes on the enveloping algebra of the $\rm SU(N)$ algebra only. In that case, the  equivalence relation cannot be applied since we lack an intrinsic formulation of the quantum theory in terms of noncommutative fields. However, the results presented in this paper indicate that their current definition in terms of ordinary fields by using the Seiberg-Witten map is a sensible one, provided one uses the $\theta$-exact Seiberg-Witten map.
\\
\begin{acknowledgments}
The work by C.P. Martin has been financially supported in part by the Spanish MINECO through Grant No. FPA2014-54154-P. The work  of J.T. is conducted under the European Commission and the Croatian Ministry of Science, Education and Sports Co-Financing Agreement No. 291823, and he acknowledges project financing by the Marie Curie FP7-PEOPLE-2011-COFUND program NEWFELPRO: Grant Agreement No. 69. 
J.Y. has been fully supported by Croatian Science Foundation under Project No. IP-2014-09-9582. We acknowledge the support of the COST Action MP1405  (QSPACE).
We would like to acknowledge L. Alvarez-Gaume and P. Minkowski for fruitful discussions and the CERN Theory Division, where part of this work was conducted, for hospitality.
We would also like to thank J. Erdmenger and W. Hollik for fruitful discussions. J.Y. would like to acknowledge the Center of Theoretical Physics, College of Physical Science and Technology, Sichuan University, China, for hospitality during his visit, as well as Yan He, Xiao Liu, Hiroaki Nakajima, Bo Ning, Rakibur Rahman, Zheng Sun, Peng Wang, Houwen Wu, Haitang Yang, and Shuxuan Ying for fruitful discussions. A great deal of computation was done using MATHEMATICA 8.0~\cite{mathematica} plus the tensor algebra package xACT \cite{xAct}. Special thanks to A. Ilakovac and D. Kekez for the computer software and hardware support.
\end{acknowledgments}

\end{document}